\documentclass[aip,apl,amsmath,amssymb,reprint,superscriptaddress]{revtex4-1}
\usepackage{graphicx}
\usepackage{amsmath}
\usepackage{dcolumn}
\usepackage{bm}
\usepackage{bbold}
\usepackage{tabularx}
\usepackage{xcolor}

\bibstyle{apsrev4-1}

\begin{document}

\title{Experimental Implementation of a Large Scale Multipost Re-Entrant Array}

\author{Maxim Goryachev}
\affiliation{ARC Centre of Excellence for Engineered Quantum Systems, University of Western Australia, 35 Stirling Highway, Crawley WA 6009, Australia}

\author{Jaemo Jeong}
\affiliation{Sungkyunkwan University, 25-2 Sungkyunkwan-ro, Myeongnyun 3(sam)ga, Jongno-gu, Seoul, South Korea}

\author{Michael E. Tobar}
\email{michael.tobar@uwa.edu.au}
\affiliation{ARC Centre of Excellence for Engineered Quantum Systems, University of Western Australia, 35 Stirling Highway, Crawley WA 6009, Australia}

\date{\today}

\begin{abstract}
We demonstrate possibilities of a large scale multi-post re-entrant cavity with two case studies implemented with the same physical structure. The first demonstration implements two discrete Fabry-P{\'e}rot cavities crossing at the centre. The configuration allows the control not only of the resonance frequencies, but also a whole band gap and transmission band of frequencies between the directly excited diagonal and a higher frequency band. The second experiment demonstrates appearance of discrete Whispering Gallery Modes on a circle of re-entrant post. With the introduction of an artificial "scatterer", we demonstrate control over the doublet mode splitting.
\end{abstract}

\maketitle

Single post re-entrant cavities\cite{reen0,reen2} have found numerous applications in many areas of engineering and physics. Among engineering applications, one may also consider material science and chemistry applications where these structures have been used to probe dielectric and magnetic properties of gases, liquids and solids\cite{May:2001aa,Harrington,enhchem,Goryachev:2014aa}.
Re-entrant cavities have become a de-facto standard component in accelerator physics\cite{Ceperley:1975aa,Geng:2005aa}. Other applications include microwave transducers for gravity wave detectors\cite{tobar1993parametric,Ballantini:2006aa,Barroso:2004aa} and dark matter detection\cite{McAllister:2016aa,Goryachev:2017aa}, hybrid quantum systems\cite{Goryachev:2014aa,Kostylev:2016ab,PhysRevB.91.140408,LeFloch16,PhysRevA.94.013812}, plasma-assisted combustion\cite{Hemawan:2009aa,Pasour:1997aa}, microwave-enhanced chemistry\cite{enhchem}, fluid-gas sensing\cite{Tsankova:2016aa,Tsankova:2017aa,Souris17}, microfluidics\cite{micf,Wang:2015aa}, tuneable filters\cite{Clark18} etc. More recently, higher order re-entrant single-post cavity modes have been studied and considered for axion Dark matter experiments\cite{HOpost17}. Performance of these types of cavities in all of these applications, as well as the list of applications itself may be enriched by further generalising the concept to multiple post re-entrant cavities. Although large scale multi-post multi-resonance structures have been analysed in detail with Finite Element Modeling (FEM), their physical implementation has not been demonstrated. In this work, we close this gap by investigating a 49 element mechanically controlled re-entrant cavity array and demonstrating it performance.

A single post re-entrant cavity is a closed conducting structure with a metallic rod protruding from one cavity wall and leaving a small gap with the opposite one. The rod makes an equivalent inductance and the gap creates an equivalent capacitance. The resonance of these two quantities is highly tuneable with the gap size\cite{reen0,reen2}. This can be done either mechanically through modifying the distance between the post and the opposite wall, for example with a piezoelectric device\cite{Carvalho14,Carvalho16}, or electrically through changing electrical distance. { A very large tuning range (over 1GHz or 19\% of the resonance frequency) has been demonstrated in a re-entrant cavity using a piezoelectric transducers\cite{C.-Carvalho:2016aa}. Similar effects were also achieved using negatively biased microwave diodes in the gap between the post and the opposite wall.} Another advantage of this type of mode is that it depends mostly on the dimensions associated with the post rather than with the cavity enclosure, thus, this isolated mode can be adjusted to any frequency within the microwave and even the millimeterwave band. 

A natural generalisation of the concept described above is a transition from a single re-entrant post to multiple posts within the same enclosure arranged in 1D\cite{Goryachev:2015aa} or 2D\cite{Goryachev:2015ab} structures with endless possibilities allowing the implementation of reconfigurable photonic topological insulators\cite{Goryachev:2016aa}. Within this approach each post represent a separate resonant element, a harmonic oscillator, coupled to others via magnetic field. Thus, the total number of re-entrant resonances equals to the number of posts. With such structures it is possible to manipulate not only eigenfrequencies of the structure but also field patterns to allow low and high concentration of magnetic fields in small regions that may be useful for many applications\cite{Goryachev:2014aa,LiFe18}. Recently it was shown experimentally, that even with a large amount of resonators, finite element analysis could be implemented to understand complicated multiple-resonant structures\cite{Leandro}.

\begin{figure}
\includegraphics[width=0.6\columnwidth]{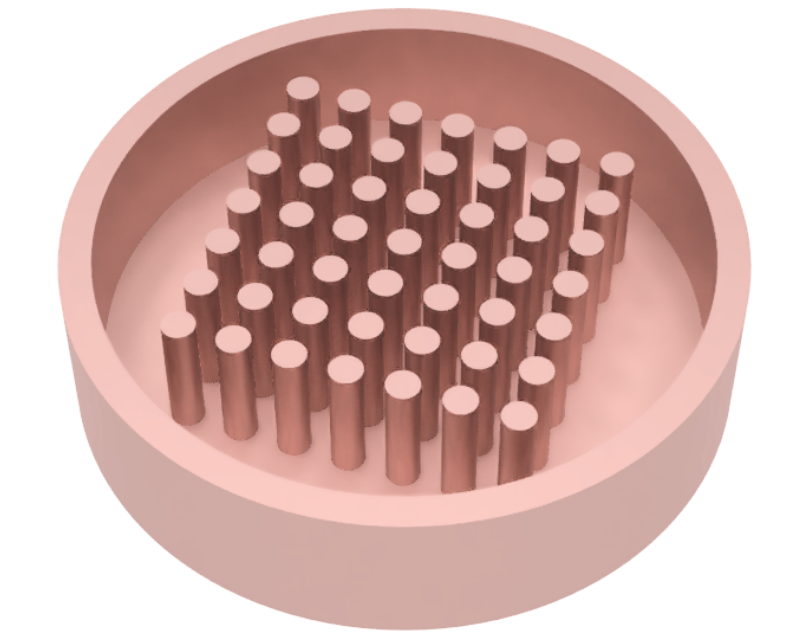}
\caption{3D picture of a 49-post re-entrant cavity without the top lid.}
\label{cavity}
\end{figure}

To demonstrate possibilities of large scale multi-post cavities experimentally, we implement a 7 by 7 regular square array of posts sharing the same walls as shown in Fig.~\ref{cavity}. The cylindrical prototype with 58~mm diameter and 3.4~mm height is manufactured from copper for room temperature operation. Each post is made as a detached element with one threaded end to control the gap mechanically in the screw-like manner. This solution allows to roughly adjust or short-circuit re-entrant gaps for each post individually or remove each post completely giving us 49 independent degrees of freedom. Each post is 2.5~mm in diameter with 5~mm spacing between centres of adjacent elements. 

The cavity is exited and probed via two loop microwave probes located at the vertices of the array. The device is characterised at room temperature with a Vector Network Analyser (VNA) measuring its transmission from one port to another from 10~MHz to 18~GHz. The VNA is calibrated up to the cavity, which excludes the interference of line resonances. {Observed modes are confirmed with FEM, particularly the eigenvalue solver of the electromagnetics module of COMSOL Multiphysics\cite{comsol}. The cavity is modelled as an empty (vacuum) 3D object bounded by a perfect conductor, thus, giving only real parts of egienvalues\cite{Goryachev:2015aa,Goryachev:2015ab}. Precision of the FEM results was controlled with the maximum size of mesh elements: this size was decreased until the calculated resonance frequency stopped changing significantly. Usually the required mesh size is determined by the gap height as the smallest feature in the design.}

\begin{figure}
\includegraphics[width=0.95\columnwidth]{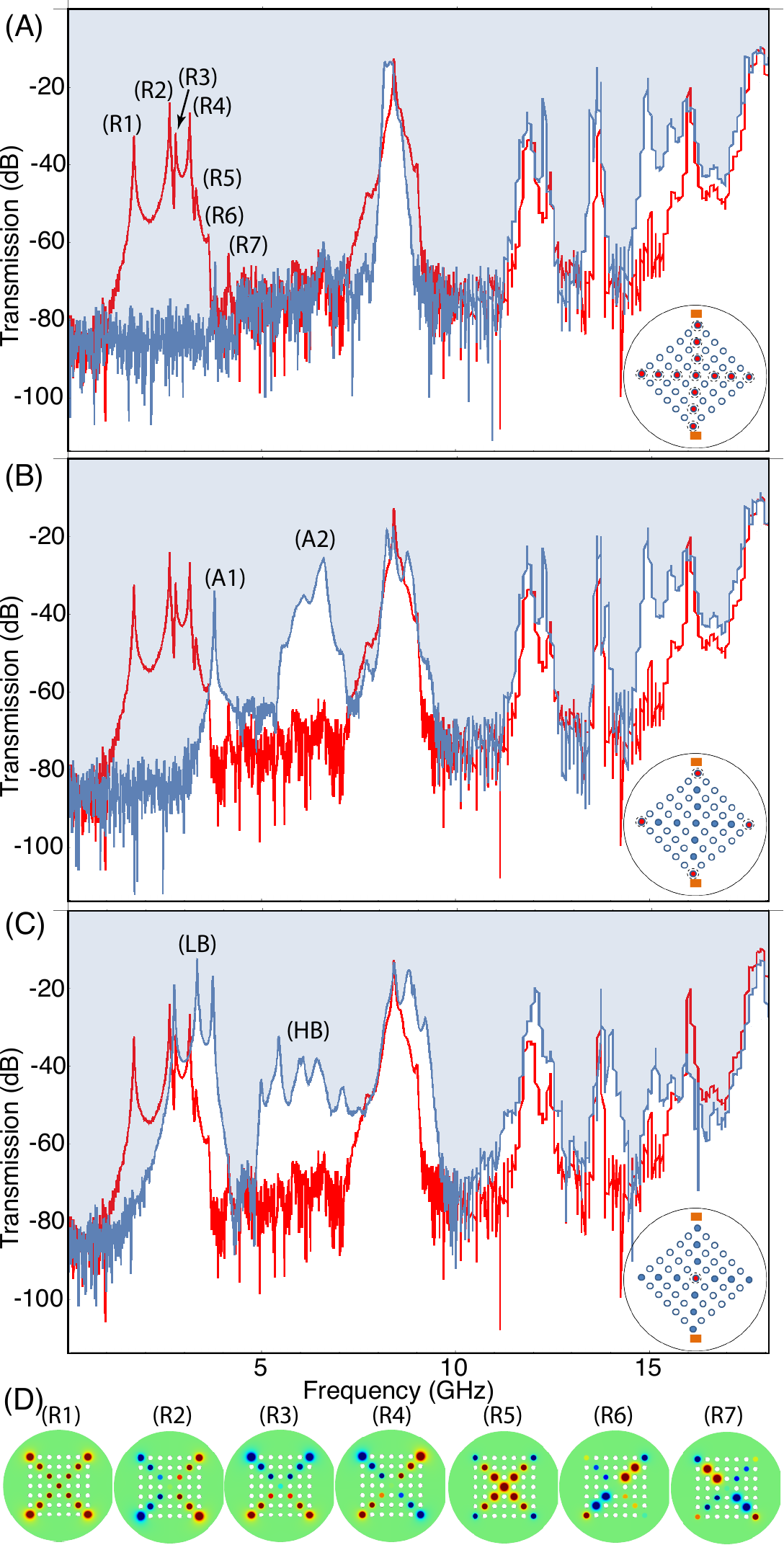}
\caption{Transmission through the cavity with the double diagonal configuration: unshaded (red) curves are the original setup, shaded (blue) curves are situation in which control post gaps are reduced: (A) control via all posts, (B) control via four vertices, (C) control via the central post. Corresponding diagrams are shown in the right bottom corner inserts: probes are shown with rectangular boxes, shaded posts participate are used in experiment, doubly circulated posts are used as control. (D) Electric field distribution under posts for seven first resonances of the original setup.}
\label{trace1}
\end{figure}

For the first experiment, we investigate the X-structure where only the posts on both lattice diagonals are involved into the mode formation, while other posts are completely removed leaving only 13 control parameters. The structure may be understood as a discrete implementation of two Fabry-P{\'e}rot 1D cavities crossing at the centre.
The system response { used further as an original system for comparison} is measured with all posts being at the same position with approximately 0.2~mm gap. This trace is shown in all three axes in Fig.~\ref{trace1} as an unshaded (red) curve. The {original} system response demonstrates six distinct resonances {(R1-R6)} in the first high transmission bad with one mode (R6) being on the higher frequency edge of the band at about 4~GHz. Another suppressed resonance (R7) is observed in the band gap stretching up to 8~GHz. {Electric field distributions in the plane of post gaps for the first seven resonances of an idealised original X-structure are shown in Fig.~\ref{trace1}. These mode structures are calculated for an ideal situation of equal gaps under each post. Note that red and blue in these plots imply opposite phases of the field under posts at the same instance of time.}

In the first comparative experiment, we varied the gaps of all posts involved simultaneously. It was observed that by increasing the gap size all seven frequencies move towards the higher frequency band eventually closing the band gap. At the same time a few other distinct resonances appear. By decreasing the gaps from the initial position all resonances are shifted to lower frequencies and reduce their absolute transmission. An extreme case of this change is shown in Fig.~\ref{trace1} (A) where the whole low frequency band disappears. The upper band of the transmission around 8~GHz reduces in size with some loss in mode structure. The leftover band gap transmission is associated with the spurious cross talk through the resonance.

For the second comparative experiment, we control only the four vertex posts. The result is shown in Fig.~\ref{trace1} (B) where only a single mode from the first band (A1) survives and moves towards the higher band. The band gap size is reduced with an additional structure (A2) appearing between 6 and 7~GHz. { Disappearance of the most of the resonances from the (R1)-(R7) group can be explained by decreased coupling to the central effective Fabry-P{\'e}rot 1D cavity because two control posts on the vertical line of posts alter coupling between these resonances and the input/output probes.}

Finally, only the central post is used to control the transmission structure of the device in the third comparative experiment. With this method, it is demonstrated that the band gap may be effectively controlled by shifting the lower transmission band (LB) towards the higher frequencies and higher transmission band (HB) towards lower frequencies as shown in Fig.~\ref{trace1} (C). { Such behaviour could be explained by the increased coupling between two orthogonal effective Fabry-P{\'e}rot 1D cavities due to the central control post serving as a scattering element.} Other control methods are also possible, for example, good results were achieved with controlling two additional middle posts for each diagonal.

It has to be noted that the higher harmonics of the re-entrant modes (more field variations along posts or in the area under posts) are usually in the higher frequency regions, { from above 10GHz in the present experiment}. These modes can be observed both from the modelling and in the experiment and thus can be excluded from this study. So, the results presented in this study can be identified as genuine first order (in the post) re-entrant modes. 


Appearance of discrete Whispering Gallery Modes (WGMs) in circular re-entrant post arrays have been previously observed with FEM methods\cite{Goryachev:2015ab}. Here, we demonstrate the modes experimentally using the same physical structure. The circular resonator is emulated using the octagon with each side consisting of 3 posts and fitted into the $7\times7$ post array. Each post constituting the octagon has a gap of $0.25$~mm.

\begin{figure}
\includegraphics[width=0.95\columnwidth]{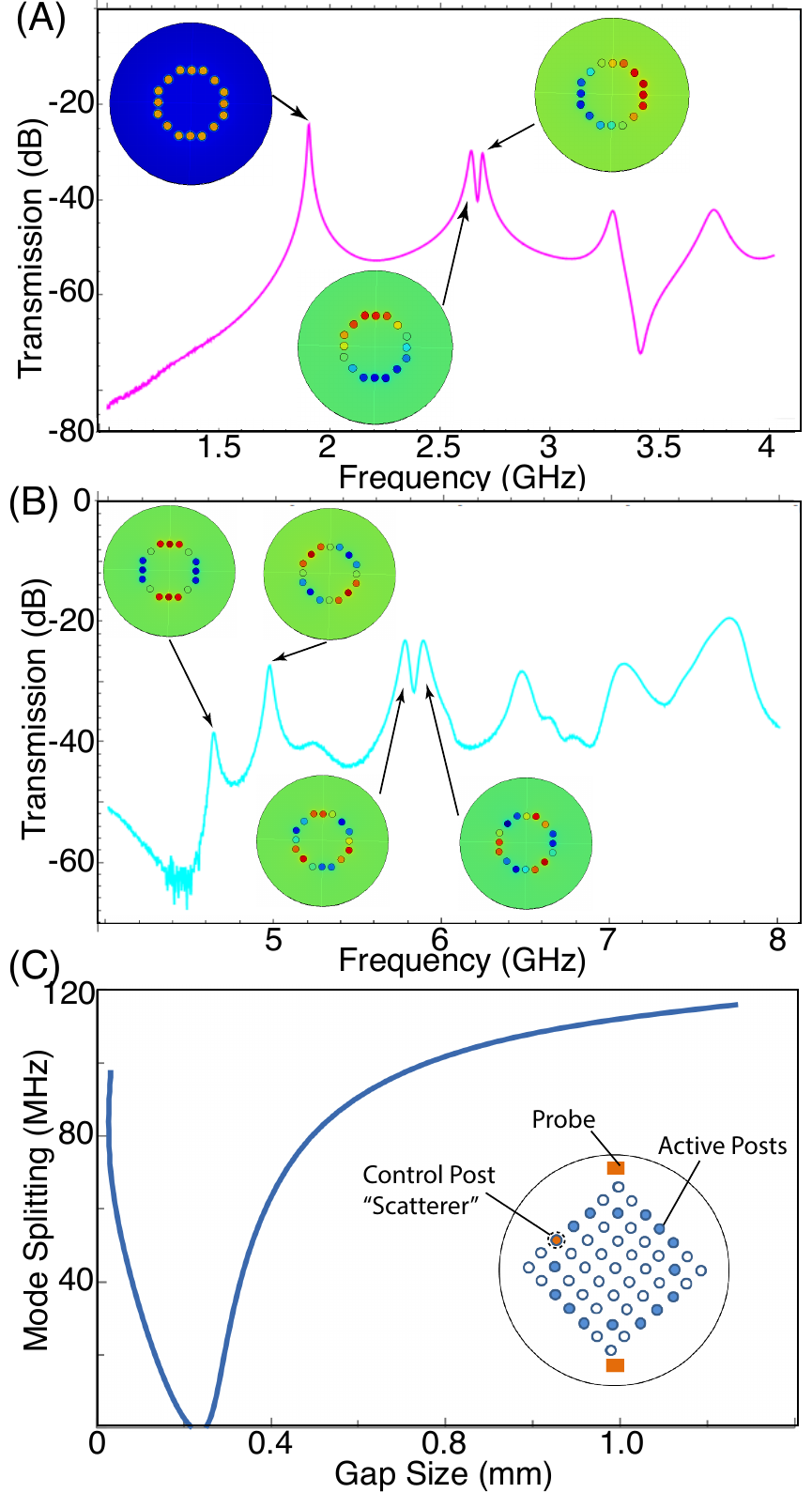}
\caption{Emulation of WHMs with a quasi-circle: (A) zeroth and first order modes, (B) second and third order modes, (C) controlling doublet splitting with an artificial "scatterer". Insets show distributions of electrical field for each mode in the gap plane in arbitrary units. }
\label{trace2}
\end{figure}

Transmission through the described system is shown in Fig.~\ref{trace2} (A) and (B). The first plot demonstrates one zeroth order and two first order modes as identified with FEM, whereas the second plot shows two second and two third order modes. Usually it is very challenging to predict exact resonance frequencies of re-entrant cavities due to their strong dependence on the gap size which is hard to measure and control. On the other hand, mutual relationships of these frequencies may be predicted and the order of modes is always determined by mutual position of the post\cite{Goryachev:2015aa, Goryachev:2015ab}. These mode structures and frequency relationships are compared to the measurements in this study. Here the mode order is determined as a number of wavelengths in the electrical field distribution along the circumference of the structure. These mode distributions are shown in insets of both plots, { where colors indicate strength and direction of the electric field in the gap plane. Thus, red and blue circles show posts with opposite directions of the field in the same instance of time.} 

A distinct feature of circular resonant structures is existence of so-called mode doublets, i.e. modes with $\pi/2$ rotational symmetry. These modes that are degenerate for ideally circularly symmetric structures form doublet in real resonators that have either anisotropy, scatterers or magnetically active elements. As seen from Fig.~\ref{trace2}, such doublets are observed for first, second and third order modes. To investigate this property, we introduces an artificial "scatterer", a post whose gap is different from others, and measured the splitting between peaks of the first order mode. The resulting mode splitting as a function of the gap size is shown in Fig.~\ref{trace2} (C). As expected splitting gets to zero almost near "normal" gap size of $0.25$~mm. At this point two modes of the doublet change their mutual positions, thus, the curve bounces from the zero line. { Also, it can be seen from Fig.~\ref{trace2} (A) and (B) that higher frequency resonances exhibit lower bandwidths. This can be explained by the fact that microwave losses increase with the frequency following the surface resistance. Moreover, the geometric factor, i.e. the ratio of energy stored in surface currents (primer source of losses) to the total cavity energy, typically increases for higher order modes leading to wider linewidths.}


In conclusion, with a 49 element re-entrant cavity, we demonstrated versatility of this type of microwave resonant structure. The same structure is used to demonstrate properties of fundamentally different resonant cavities: a double Fabry-P{\'e}rot-like system and a WGM resonator. In both cases, we observe and are able to control some important features of these systems: a band gap and a doublet mode splitting. Agreement between experimental observations and FEM confirms previous theoretical investigations of large scale multipost re-entrant cavities \cite{Goryachev:2015aa,Goryachev:2015ab,Goryachev:2016aa}. We demonstrate that such cavities are robust, highly tunable, versatile whose performance may be easily engineered. This demonstration opens a wide range of new applications from tunable filters to dark matter detection and hybrid quantum systems. { The versatility of such many-post structure could be achieved by implementing an {\it in-situ} tuning mechanism either a piezoelectric transducer or a microwave diode\cite{C.-Carvalho:2016aa}.}


This work was supported by the Australian Research Council grant number CE170100009. The authors also thank Leandro de Paula for his initial design work on the resonant structure reported in the work.

\section*{References}

%

\end{document}